\documentclass[numbers]{elsarticle}
\usepackage{latexsym}
\usepackage{natbib}
\usepackage[utf8]{inputenc}
\usepackage[hidelinks]{hyperref}
\usepackage{longtable, lineno}

\makeatletter
\def\ps@pprintTitle{%
	\let\@oddhead\@empty
	\let\@evenhead\@empty
	\def\@oddfoot{\centerline{\thepage}}%
	\let\@evenfoot\@oddfoot}
\makeatother

\usepackage{longtable}
\usepackage{graphicx}
\usepackage{epstopdf}
\usepackage{float}
\usepackage{xcolor}

\newcommand\sq[1]{\fboxsep=2mm \fboxrule=0.2mm \fcolorbox{black}{#1}{\raisebox{0.7ex}{\null}}}
\definecolor{aqua}{rgb}{0.0, 1.0, 1.0}

\begin{document}

	\title{Mapping Correlations of Psychological and Connectomical Properties of the Dataset of the Human Connectome Project with the Maximum Spanning Tree Method}
	
	
	\author[p]{Balázs Szalkai\corref{cor1}}
	\ead{szalkai@pitgroup.org}
	\author[p]{Bálint Varga}
	\ead{balorkany@pitgroup.org}
	\author[p,u]{Vince Grolmusz\corref{cor1}}
	\ead{grolmusz@pitgroup.org}
	\cortext[cor1]{Joint corresponding authors}
	\address[p]{PIT Bioinformatics Group, Eötvös University, H-1117 Budapest, Hungary}
	\address[u]{Uratim Ltd., H-1118 Budapest, Hungary}

	\date{}

\begin{abstract}
	We analyzed correlations between more than 700 psychological-, anatomical- and connectome--properties, originated from the Human Connectome Project's (HCP) 500-subject dataset. Apart from numerous natural correlations, which describe parameters computable or approximable from one another, we have discovered numerous significant correlations in the dataset, never described before. We also have found correlations described very recently independently from the HCP-dataset: e.g.,  between gambling behavior and the number of the connections leaving the insula.
\end{abstract}

\maketitle

\section{Introduction}

A large amount of human psychological and behavioral data were collected and deposited in the last several decades worldwide. In the framework of the Human Connectome Project \cite{McNab2013} those type of  data were enriched with structural and functional MR images of the same subjects. In the present contribution, we are analyzing the graph-theoretical properties of the connectomes as well as the psychological and behavioral test data that were published in the Human Connectome Project's \cite{McNab2013} anonymized 500 Subjects Release. Our goal is finding correlations between those highly inhomogeneous data items (containing graph properties, psychological test scores, brain area volumes, etc.). We are considering Pearson's product-moment correlation, which well describes the linear connections between attributes, and also Spearman's rank correlation, which well describes non-linear connections between attributes \cite{Wonnacott1972}.

Some of the strongest correlations are natural, describing closely related quantities (e.g., the volume and the relative volume of the same brain area, or graph maximum matching numbers and minimum vertex covers). Some of them are novel, and are detailed in this contribution, and some of them were discovered in the recent years (e.g., the connection between gambling behavior and the number of connections crossing the insular cortex \cite{Dymond2014}).

\subsection{Maximum Weight Spanning Trees of Correlations}

Suppose we have a large number of attributes, describing the properties of a complex system. Frequently, a straightforward step in their analysis is the computation of the pairwise correlations of the attributes. If we have $n$ attributes, then one can form $n(n-1)/2$ pairs from them, so we will need to compute that many pairwise correlations as well. Identifying the most ``important'' correlations from the set is, generally, not an easy task. 

A possible natural requirement is generating a ``non-redundant set'' of correlations in the following sense: Suppose that random variable A strongly correlates with B, and random variable B strongly correlates with C, then, usually, A and C are also strongly correlated. Now, if we want to find a non-redundant set of correlations between A, B and C, it is an obvious idea to choose the two strongest correlation between them, say between A and B and between B and C, and to leave out the weakest, say the one between C and A. We can visualize those non-redundant ``strong''' or ``important'' correlations by a graph on vertices A, B and C, with two edges: AB and BC.

In the general case when $n$ attributes are considered, we are interested in cycle-free connected graphs with the highest possible total weight of edges, where the weight of an edge is defined as the absolute value of the corresponding correlation. The cycle-free property ensures the non-redundancy, and the highest total weight of the absolute values of the correlations ensures that we have chosen the ``most relevant'' ones. 

The idea of constructing the maximum weight spanning tree from the pairwise correlation coefficients was applied before us in several settings.

Mantegna \cite{Mantegna1999} constructed a graph from financial equities, traded in stock markets, and weighted the edges of the graph by the correlations computed from the time series of the logarithms of the stock prices. It was found that -- essentially -- a maximum-weight spanning tree well-describes several known relations and suggests numerous new ones between the time series. In \cite{Bouchaud2003} Section 9.3.5 and \cite{Heimo2009} the maximum-weight spanning trees are computed explicitly in similar settings.

In \cite{Zivkovic2009} correlations related to the co-expression of gene-pairs of the yeast ({\em S. cerevisiae}) were computed, and a graph was constructed with the genes as the vertices and the co-expression correlation-weighted edges for each pair of genes. Next, a maximum weight spanning tree was computed and visualized for demonstrating a non-redundant system of strong correlations between the genes. (\cite{Zivkovic2009}, Fig. 6).

More recently, in \cite{Ha2015}, the maximum-weight spanning tree of the correlations is applied for feature selection in weakly-structured multimedia data. In \cite{Brida2015} a similar method is used for finding related attributes in a regional Italian hospitality industry.

\section{Materials and Methods}

Our data source is the Human Connectome Project's \cite{McNab2013} anonymized 500 Subjects Release. In the dataset diffusion- and functional MRI data, psychological test results, and some cognitive data are made public. Here we list the different types of data applied by us.

The subject data table from the Human Connectome Project consists of 527 rows (corresponding to subjects) and 451 attributes. 

Diffusion MRI images of the subjects were processed by the researchers of the Human Connectome Project with the Freesurfer software suite \cite{Fischl2012} to obtain the size of various regions of interests (ROIs), i.e., the area, thickness and volume of major cortical and sub-cortical areas of the brain. For cortical regions, only average thickness and surface area were available, so we multiplied these two quantities to obtain the approximate volume of the ROI. We divided the volume of an ROI by \texttt{FS\_Mask\_Vol} (Freesurfer Brain Mask Volume, i.e., roughly the brain volume) to obtain the {\em relative volume} of that region. We intended to compensate for brain size because it is already well known that males on the average have larger brains than females \cite{Witelson2006}. We added these new, relativized attributes to the data table.

Several psychological and cognitive tests were also performed with the subjects. These included the MMSE (Mini Mental State Examination), various NIH Toolbox \cite{Weintraub2013} cognitive tests (Picture Sequence Memory Test, Dimensional Change Card Sort Test, Flanker Inhibitory Control and Attention Test, Oral Reading Recognition Test, Picture Vocabulary Test, Pattern Comparison Processing Speed Test, List Sorting Working Memory Test), NIH Toolbox Emotion Domain (Anger-Affect, Anger-Hostility, Anger-Physical Aggression, Fear-Affect, Fear-Somatic Arousal, Sadness, General Life Satisfaction, Meaning and Purpose, Positive Affect, Friendship, Loneliness, Perceived Hostility, Perceived Rejection, Emotional Support, Instrumental Support, Perceived Stress, Self-Efficacy), a test for self-regulation/impulsivity (Delay Discounting), Penn Line Orientation Test, Penn Continuous Performance Test, Penn Word Memory Test and Penn Emotion Recognition Test.

The subjects were also asked to perform some fMRI tasks, including identifying random and non-random shape movements, a working memory test (places, faces, body parts, and tools), language, math and gambling tasks. 

We also added numerous attributes to the data table corresponding to various graph parameters of the connectomes of the subjects. These included the total number of the connectome edges, counted with weights, maximum matching and minimum vertex cover, Hoffman's bound (a bound for the chromatic number), the eigengap of the transition matrix (which is a quantity connected with the properties of random walks on the graph), and the total number of edges exiting different lobes of the brain. These graph-theoretical quantities of the connectomes were defined in details and also computed in the articles \cite{Szalkai2015, Szalkai2015c}.

We calculated the correlations between all possible pairs of the attributes (columns of the database). The goal was to obtain a non-redundant list of important correlations. Our observation was that correlation is transitive in most of the cases, so if A correlates with B and B correlates with C, then this usually implies that A correlates with C in some degree. Therefore, if we consider a complete graph whose vertices are the attributes, and whose edges represent correlations between two attributes, then our goal can be reformulated as follows: we want to find a subgraph without cycles (because cycles usually mean redundant correlations), and whose edges correspond to relatively large correlations (because larger correlations are more important than the others).

This optimization problem is essentially a maximum weight spanning tree problem, which can be solved by graph theoretical algorithms such as the Kruskal algorithm \cite{Lawler1976}. The classical question is finding a {\it minimum} weight spanning tree, but by a straightforward transformation, the algorithm for the minimum weight spanning tree can be used for finding the maximum weight spanning tree. Indeed, let $w_e$ be the weight (or in our specific application, the correlation) corresponding to edge $e$, and let $W=\max w_e$, taken for all edges $e$. Then the maximum-weight spanning tree with weights $w_e$ is, at the same time, the minimum-weight spanning tree with weights $W-w_e$.

A similar correlation-based maximum spanning tree approach has already been used by other researchers \cite{Mantegna1999,Bouchaud2003,Heimo2009,Zivkovic2009,Ha2015,Brida2015}. We applied this method to the HCP (Human Connectome Project) subject data table in its anonymized 500 Subjects Release \cite{McNab2013}: \url{http://www.humanconnectome.org/documentation/S500} in order to search for connections between psychological and cognitive scores, demographic data, and brain ROI sizes.

The possible age groups of the subjects were 22-25, 26-30, 31-35 and 36+. Only 3 of the subjects were 36 years or older. We translated the age groups to numbers the following way: 0 meant 22-25, 1 meant 26-30, 2 meant 31-35 and 3 meant 36+. We translated the ``gender'' attribute to 1 (male) and 2 (female). We had to convert these attributes to numbers so we can calculate correlations between them and other attributes.

We have computed both the Pearson's product-moment correlation (this is ``the correlation'', most frequently computed in science), which well describes the linear connections between attributes, and also Spearman's rank correlation, which well describes non-linear connections between attributes \cite{Wonnacott1972}.

\section{Results}

\subsection*{Maximum spanning tree of Pearson's correlations}

The maximum spanning tree is visualized on Figure 1 and in a more viewable form in an interactive figure at \url{http://pitgroup.org/static/graphmlviewer/index.html?src=correl_spanning_tree.graphml}.

The spanning tree had 716 edges with non-zero correlation, and 717 attributes, so the graph contained 717 vertices. The weakest edge still had 15\% correlation. The complete table describing the maximum-weight spanning tree, where the weights are the correlations, is given as Supporting Table S1.

The significance of the correlations was determined by multiplying their p-value with the total number of edges in the graph, which was $717*716/2 = 256,686$, because we wanted to correct for multiple observations: we made as many observations as the number of edges in the graph. Thus, we obtained an upper limit of the p-value of the correlations. This meant that eight correlations have been deemed insignificant, but all the other edges of the spanning tree belonged to significant correlations (this meant $716-8=708$ edges). This indicated that almost all attributes of the database are more or less interdependent of each other.

\begin{figure}
\centering
\includegraphics[width=12cm,keepaspectratio]{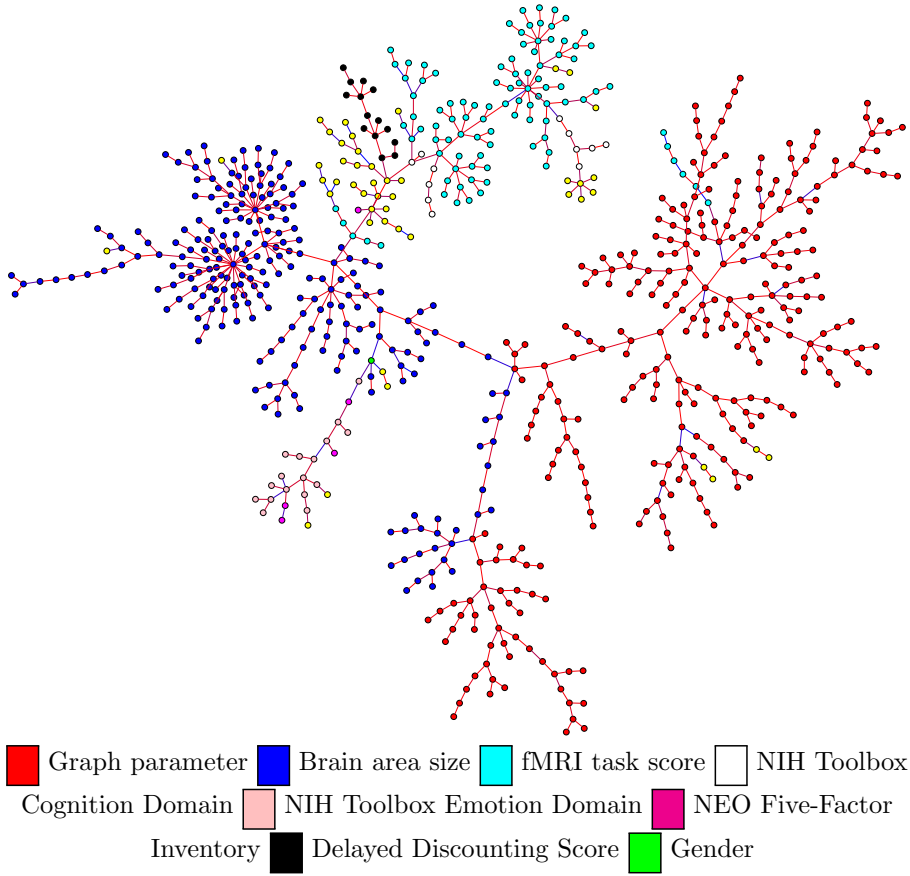}
\sq{red} Graph parameter \sq{blue} Brain area size \sq{aqua} fMRI task score
\sq{white} NIH Toolbox Cognition Domain \sq{pink} NIH Toolbox Emotion Domain \sq{magenta} NEO Five-Factor Inventory 
\sq{black} Delayed Discounting Score \sq{green} Gender
\caption{The maximum-weight spanning tree of the correlations of 717 quantities. The tabular data is given as supporting Table S1, and the interactive version of this figure can be viewed with node-labels at \url{http://pitgroup.org/static/graphmlviewer/index.html?src=correl_spanning_tree.graphml}.}
\end{figure}

The supplementary Table S1 contain a list of all the correlations in the spanning tree. 185 correlations referred to attributes which are either completely dependent on each other or are very close to a linear dependence (over 90\% positive or negative correlation). Gray matter volume correlated with the volume of many cortical regions, which is not very surprising since these regions comprise the cortical gray matter. These included the superior frontal gyrus, the lateral orbitofrontal cortex, the precuneus, the middle temporal gyrus, the precentral gyrus, the fusiform gyrus and the rostral middle frontal gyrus (the list is incomplete).

Elements of the NIH Toolbox Emotion Domain showed a strong correlation with each other: sadness and fear affect, sadness and anger affect, sadness and perceived stress, sadness and loneliness, loneliness and perceived rejection, perceived hostility and perceived rejection, friendship and emotional support, friendship and loneliness (negative), life satisfaction and meaning and purpose, emotional and instrumental support, life satisfaction and positive affect, anger-hostility and perceived stress, life satisfaction and perceived stress (negative), perceived stress and self-efficacy (negative), fear affect and fear-somatic arousal. Even the weakest of these correlations was 49\% strong, the strongest (sadness and fear affect) being 72\% strong.

There were 17 attributes in the NIH Toolbox Emotion Domain, and they almost represented a connected subgraph in the spanning tree (see Figure 2). This means that, by including the NEO-FFI Agreeableness attribute (which correlates positively with NIH Emotional Support and negatively with NIH Anger-Physical Aggression), we get an 18-vertex set which spans a 17-edge subtree of the spanning tree. This means that these attributes are strongly correlated with each other, comprising an important correlated subset of all the attributes.

\begin{figure}
\centering \includegraphics[width=12cm,keepaspectratio]{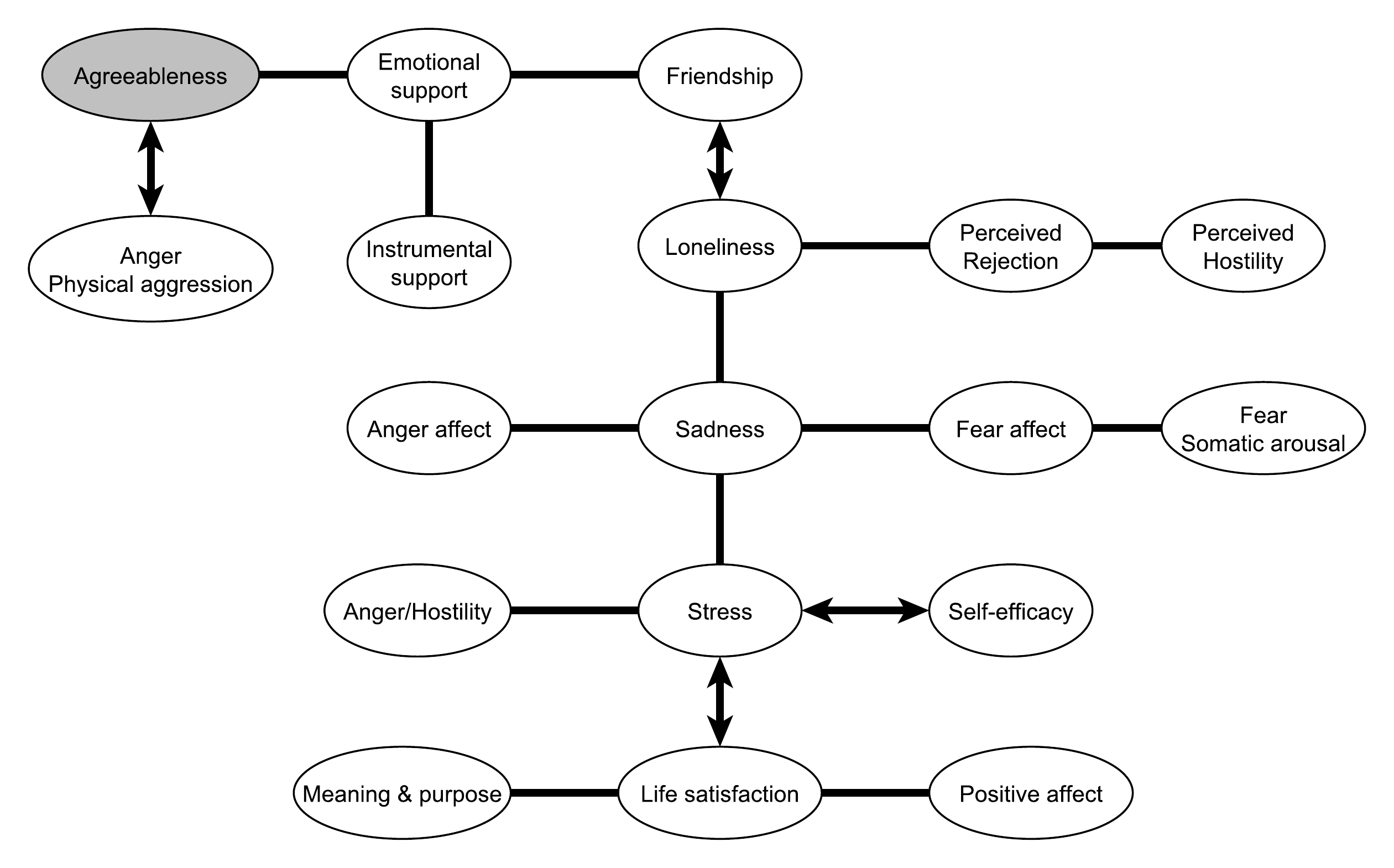}
\caption{The attributes of the NIH Toolbox Emotion Domain. The connections are significant correlations in the constructed spanning tree.}
\end{figure}

The NEO-FFI personality scores appear to be correlated with certain NIH toolbox items, in a sense that they are leaves of those NIH toolbox items in the tree. NEO-FFI Neuroticism is a leaf of NIH Perceived Stress (correlation: 70\%). NEO-FFI Conscientiousness is a leaf of NEO-FFI Neuroticism (correlation: -41\%). NEO-FFI Extraversion is a leaf of NIH Friendship (correlation: 50\%). An exception is NEO-FFI Agreeableness, which is not a leaf as it is connected to both NIH Anger-Physical Aggression (correlation: -45\%) and NIH Emotional Support (correlation: 35\%). An interesting finding is that NEO-FFI Openness to Experience is a leaf of the NIH Toolbox Oral Reading Recognition Test Unadjusted Scale Score (correlation: 32\%). This indicates that good reading skills and openness to experience frequently come together. To sum up, we can observe a strong connection between the NEO-FFI personality scores and the NIH toolbox positive or negative affect scores. This can mean multiple things. Our interpretation is that our personality defines our emotions to a great extent, which can be measured by statistical means.

The NIH Toolbox Oral Reading Recognition Test is a very important hub in the spanning tree. Its unadjusted and age-adjusted versions are correlated with: Picture vocabulary task age-adjusted score (71\%), Language task accuracy (49\%), Penn Progressive Matrices Number of Correct Responses (47\%), NIH Toolbox 2-minute Walk Endurance Test Age-Adjusted Score (43\%), Delay Discounting Subjective Value for \$40K at 1 year (35\%), Short Penn Continuous Performance Test Specificity (33\%), NEO-FFI Openness to Experience (32\%), MMSE score (31\%) and Penn Word Memory Test Total Number of Correct Responses (31\%). Most of these are cognitive tests. The walk endurance test score is very interesting since it draws a connection between physical and intellectual fitness. It also seems that people, scoring higher on the oral reading test, value a delayed payment higher than those with lower scores. There can be multiple reasons for this; the observed correlation could follow from their possible better financial status, financial skills or greater patience. We have covered the NEO-FFI connection above.

It seems that gender (male=1, female=2) correlates with the NIH Toolbox Grip Strength Test Age-Adjusted Scale Score (-75\%), the brain mask volume (-67\%), the optic chiasm volume (-41\%) and the NIH Toolbox Anger-Physical Aggression Survey Unadjusted Scale Score (-26\%). All of these are significant. This means that men, on average, have greater grip strength, brain volume (especially optic chiasm volume) and are more aggressive physically.

The score of the walk endurance test correlates with the taste intensity score, although with one of the smallest significant observed correlations (-25\%). Walk endurance seems to correlate with less perceived bitterness of quinine. The cause of this correlation is unknown to us.

Another interesting correlation is NIH Toolbox 9-hole Pegboard Dexterity Test: Age-Adjusted Scale Score, versus Maximum matching (left hemisphere, weight function: mean fractional anisotropy) (27\%). The corrected p-value is rather high when compared to the other correlations (p=0.003) but still significant. This is an important correlation because it shows a significant relationship between a parameter of the brain graph (connectome) and a brain performance score.

We expected the Age attribute to be a major hub of the spanning tree, but, interestingly, it was only a leaf of the attribute Right hemisphere cortical gray matter relative volume (correlation: -23\%, p=0.037). We think this is because all the subjects were rather young, so their cognitive and psychological scores did not depend heavily on their age. Still, it is interesting that cortical gray matter relative volume correlates with age even among relatively young subjects.

\subsection*{Maximum spanning tree of Spearman's rank correlations}

We also investigated how the spanning tree changes if we use Spearman's rank correlation coefficient instead of the classical (Pearson's) correlation coefficient  \cite{Wonnacott1972}. We computed the rank correlation of two attributes as follows: the values were ordered independently for the two attributes, and then each value was substituted with its rank in the ordering. If two or more attributes were equal, then their rank was defined as the same number, and the succeeding rank(s) were omitted. In other words, the rank of a value $v$ was defined as the number of values strictly smaller than $v$, plus one. For example, the values $1, 1, 4, 13, 13, 45$ were assigned the ranks $1, 1, 3, 4, 4, 6$.

We calculated a maximum weight spanning tree for the rank correlations as well. The detailed data of the tree is given in Table S2 in the supporting material. The interactive figure with vertex labels is available at the address \url{http://pitgroup.org/static/graphmlviewer/index.html?src=correl_spanning_tree_rank.graphml}

This tree will be referred to as the {\em rank spanning tree} from now on, and the one using the traditional correlation coefficient will be referred to as the {\em original spanning tree}.

We examined those edges which are present in exactly one of the spanning trees. In the following analysis, we omitted the edges concerning two graph parameters or two brain area sizes. We also omitted those that connect two nodes (attributes) that can be exactly calculated from each other (e.g. the number of false positives and true negatives for some test). Edges which connect two attributes referring to subscores of the same task are also omitted.

There were $98$ edges in each spanning tree that were not present in the other tree. As the trees contained $716$ edges each, this means that approximately $13.7\%$ of the edges were unique to the containing tree. In other words, the two trees were rather similar, having $86.3\%$ of the edges in common. Among the unique edges, the vast majority were omitted from the analysis, or connected very similar nodes. For example, the unadjusted and age-adjusted version of an attribute, or the median reaction time and the number of correct responses for a task, or scores for two closely related tasks were considered similar attributes.

Some cognitive attributes were connected in a different way in the rank correlation tree when compared to the original correlation tree. For example, the IWRD and MMSE total scores were connected to the English reading score in the original tree, but connected to the Picture Vocabulary score in the rank tree. The computed score for the NIH Toolbox Words-In-Noise test was connected to a sub-score of the Social fMRI task in the original tree, but connected to, again, the Picture Vocabulary score in the rank tree. This may suggest that the Picture Vocabulary score is strongly connected to these complex cognitive scores, but not necessarily in a linear fashion. The same goes for the Working Memory fMRI task, which was connected to the Picture Vocabulary score in the original tree, but connected to both a sub-score of the Relational task and the number of correct responses in the Penn Matrix Test in the rank tree.

An interesting connection between the volume of the right lateral ventricle and the number of correct happy identifications in the Penn Emotion Recognition Test was included in the rank correlation tree. However, we should note that the corrected p-value for this edge was very large, about $72$. This means that there is a large likelihood that this connection was included by mere chance.

Regarding the graph parameters and the brain ROI sizes, the volume of the anterior corpus callosum and the relative volume of the mid-posterior corpus callosum were connected to two graph parameters (sum and minimum cut) in the original tree, but, surprisingly, these natural connections were no longer present in the rank tree. A graph parameter related to eigenvalues (Graph\_Left\_AdjLMaxDivD\_FiberNDivLength) was connected to the NIH Toolbox Odor Identification Test unadjusted score in the normal tree, but this connection was not included in the rank tree, which, on the other hand, contained an edge between Age and the age-adjusted score of the odor identification test (corrected p-value: $6\%$. That connection is somewhat surprising since the age-adjusted scores should not be correlated with age. This could mean that some tests are not well adjusted for age. Another possible explanation could be that, although the test score does not change significantly for the same person over the person's lifetime, but different generations may have different mean scores due to environmental factors.

\section{Conclusions} We have analyzed both the original Pearson's and Spearman's rank correlations with 717 psychological, anatomical and connectome-properties originated from the Human Connectome Project's subject 500-release. Apart from numerous natural correlations that describe parameters computable or approximable from one another, we have discovered numerous strong, significant correlations in the dataset, never described before. 

\section*{Data availability} The MRI and the behavioral and demographic data the Human Connectome Project (HCP) can be accessed at \url{https://db.humanconnectome.org/}. The braingraphs, computed by us from the HCP data is available at the site \url{http://braingraph.org/download-pit-group-connectomes/}, without any registration. 

\section*{Acknowledgments}
Data were provided in part by the Human Connectome Project, WU-Minn Consortium (Principal Investigators: David Van Essen and Kamil Ugurbil; 1U54MH091657) funded by the 16 NIH Institutes and Centers that support the NIH Blueprint for Neuroscience Research; and by the McDonnell Center for Systems Neuroscience at Washington University.



\end{document}